\documentclass[aip,jcp]{revtex4-1}

\usepackage{amssymb}
\usepackage{graphicx}
\usepackage{dcolumn}
\usepackage{bm}
\usepackage{longtable}

\begin{document}

\newcommand{\cm}{cm$^{-1}$} \newcommand{\A}{\AA$^{-1}$} \newcommand{\Q}{\mathbf{Q}}
\newcommand{\PP}{\mathcal{P}}

\draft

\title{A neutron diffraction study from 6 to 293 K and a macroscopic-scale quantum theory of the hydrogen bonded dimers in the crystal of benzoic acid }

\author{Fran\c{c}ois Fillaux}
\email{francois.fillaux@upmc.fr}  \homepage{http://ulysse.glvt-cnrs.fr/ladir/pagefillaux.htm}
\affiliation{LADIR-CNRS, UMR 7075 Universit\'{e}
 Pierre et Marie Curie, 2 rue Henri Dunant, 94320 Thiais, France}
\author{Alain Cousson}
\affiliation{Laboratoire L\'{e}on Brillouin (CEA-CNRS), C.E. Saclay, 91191
Gif-sur-Yvette, cedex, France}

\date{\today}

\begin{abstract}
The crystal of benzoic acid is comprised of tautomeric centrosymmetric dimers linked through bistable hydrogen bonds. Statistical disorder of the bonding protons is excluded by neutron diffraction from 6 K to 293 K. In addition to diffraction data, vibrational spectra and relaxation rates measured with solid-state-NMR and quasi-elastic neutron scattering are consistent with wave-like, rather than particle-like protons. We present a macroscopic-scale quantum theory for the bonding protons represented by a periodic lattice of fermions. The adiabatic separation, the exclusion principle, and the antisymmetry postulate yield a static lattice-state immune to decoherence. According to the theory of quantum measurements, vibrational spectroscopy and relaxometry involve realizations of decoherence-free Bloch states for nonlocal symmetry species that did not exist before the measurement. The eigen states are fully determined by three temperature-independent parameters which are effectively measured: the energy difference between tautomer sublattices; the double-well asymmetry for proton oscillators; the delocalization degree of the wavefunctions. The spontaneous decay of Bloch states accounts for relaxometry data. On the other hand, static states realized by elastic scattering account for  diffraction data. We conclude that both quantum and classical physics hold at every temperature.

\end{abstract}

\maketitle

\section{Introduction}

The crystal of benzoic acid (BA, C$_6$H$_5$COOH) \cite{HU1,*HU2,*HONM,MGE,*ME,*ST,*SMME,*ME2,*HH,*SVS,*BHJ,
*LR1,*TT,*LRB,*XHJT,*WNH,WSF1,*WSF2,*WXSF,NBMJHT,PFJHT,FLR,FRLL} and that of potassium hydrogen carbonate (KHCO$_3$) \cite{TTO1,*TTO2,Fil2,*FTP,IKSYBF,EGS,FCKeen,*FCG2,*FCG3,*FCG5,Fil7} are reference cases for thermal interconversion of cyclic dimers, namely (C$_6$H$_5$COOH)$_2$ or (HCO$_3^-)_2$, linked through bistable O--H$\cdots$O hydrogen bonds. It has been observed for a long time that the coexistence of two configurations, say ``$L$'' for $\mathrm{O1-H}\cdots \mathrm{O2}$ and ``$R$'' for $\mathrm{O1}\cdots \mathrm{H-O2}$, leads to centrosymmetric configurations, $LL$ and $RR$, exclusive of the less stable entities, $LR$ or $RL$. The ratio $RR:LL$ changes smoothly over a broad range of temperature without any structural transition. BA is traditionally regarded as a prototype for a class of chemical reactions involving proton transfer and tautomerism, that is of fundamental importance in many fields across physics, chemistry and biology. Experimental and theoretical studies are expected to enlighten the quantum behavior of hydrogen bonds in complex environments, with particular emphasis on proton tunneling and on the emergence of the classical regime at elevated temperatures. Published works can be categorized as follows: (i) quantum, for vibrational spectra; \cite{FLR,FRLL,Fil2,Fil7} (ii) semiclassical, for coarse-grained measurements that cannot resolve individual quantum levels, what is essentially the case for solid-state NMR or quasi-elastic neutron scattering (QENS); \cite{MGE,EGS} (iii) theoretical, for computed Born-Oppenheimer (BO) potentials. \cite{MTK,*UT,*MDM,DLO,*YRBBDL,SBA,*Kim,*SFS,*MKN,*Luckhaus,*MDK,*BS,*BSS,*MDJ,*SSF2,*Luckhaus2,*SDP} Not surprisingly, different approaches lead to severe conflicts of interpretation showing that there is so far no satisfactory interpretational framework.

As a matter of fact, experiments conducted with different techniques have been interpreted in contradictory ways. For example, whereas the ratio $RR:LL$ is commonly attributed to disorder, \cite{TTO1,TTO2,WSF1,WSF2} neutron diffraction and vibrational spectra provide evidences of quantum correlations excluding disorder. \cite{FCKeen,*FCG2,*FCG3,*FCG5,Fil7} Further conflicts appear when the relaxation rate measured with NMR/QENS is modelled by stochastic jumps of rigid proton pairs, induced by a thermal bath. \cite{ME,EGS} By contrast, the quantum regime on the timescale $\sim 10^{-12} - 10^{-14} s$ is attested to by vibrational spectra, from cryogenic to room temperature, \cite{FLR,FRLL} and symmetry related selection rules for Bloch states extended throughout the crystal exclude local dynamics. In addition, the tiny proton-proton coupling terms effectively observed exclude rigid pairs and the discrete density-of-states of the crystal is markedly different from an incoherent thermal bath. \cite{FLR,FRLL,Fil7} The point at issue is whether these conflicts arise because the fundamental nature of the crystal is beyond the reach of measurements, or whether the ad hoc interpretative frameworks are not appropriate to recognizing common features under the light of different techniques.

Theoretical studies of the crystal of BA with computational quantum-chemistry are rather scarce. \cite{JT,*TKII,*LZT,DLO,*YRBBDL} The most detailed works deal with the isolated dimer of formic acid (FAD). \cite{SBA,*Kim,*SFS,*MKN,*Luckhaus,*MDK,*BS,*BSS,*MDJ,*SSF2,*Luckhaus2,*SDP,MTK,*UT} Within the framework of the BO separation, it is postulated that every experiment can be represented by classical particles moving along definite trajectories. This ``local realism'' leads to the notions of ``reaction-path'', ``promoter mode'' and ``transition-state'' for ``proton-transfer'', which are strictly alien to quantum physics. \cite{SSF} In fact, such theoretical studies are hardly conclusive because the dominant mechanism is largely model-dependent: either stepwise for molecular dynamic simulations, \cite{MTK,*UT,*MDM,DLO,*YRBBDL} or concerted, of course, for ab initio calculations with rigid pairs. \cite{SBA,*Kim,*SFS,*MKN,*Luckhaus,*MDK,*BS,*BSS,*MDJ,*SSF2} Whatever the mechanism, quantum chemistry emphasizes that proton transfer should be promoted by a shortening of the O$\cdots$O bonds, what is in conflict with the adiabatic separation of protons and heavy atoms attested to by experiments. \cite{FCG5,FLR,FRLL} It is, therefore, an open question whether or not computational chemistry can capture the essential features of tautomerism.

Our purpose is to show that we can eliminate every conflict of interpretation if we adopt the radical viewpoint that the crystal is a macroscopic-scale quantum system, essentially decoherence-free, at every temperature below melting or decomposition, and on timescales ranging from vibrational spectroscopy to diffraction. \cite{FCKeen,FLR,FRLL,Fil7} Consequently, the classical behavior of nuclei and the intuitive notions forwarded by computational chemistry are abandoned. This quantum view of a macroscopic-scale open system is clearly in conflict with the theory of decoherence stipulating an irreversible transition to the classical regime due to the exceedingly fast loss of quantum coherence leaking out into the environment, at a rate growing exponentially with the number of particles. \cite{Zurek1} On the contrary, we shall show that the sublattice of bonding protons is immune to irreversible decoherence and that this immunity increases with the size of the crystal. This quantum theory was first elaborated for the crystal of KHCO$_3$ and confronted successfully with observations. \cite{Fil7,FCG5} The key point is that bonding protons behave as fermions, what is, to the best of our knowledge, unprecedented in the context of hydrogen bonding. Our purpose is to extend this theoretical framework to the more complex crystal of BA and to re-examine the conflicts of interpretation engendered by classical physics under the light of the theory of quantum measurements.

\begin{figure}[!hbtp]
\begin{center}
\includegraphics[scale=0.5, angle=0]{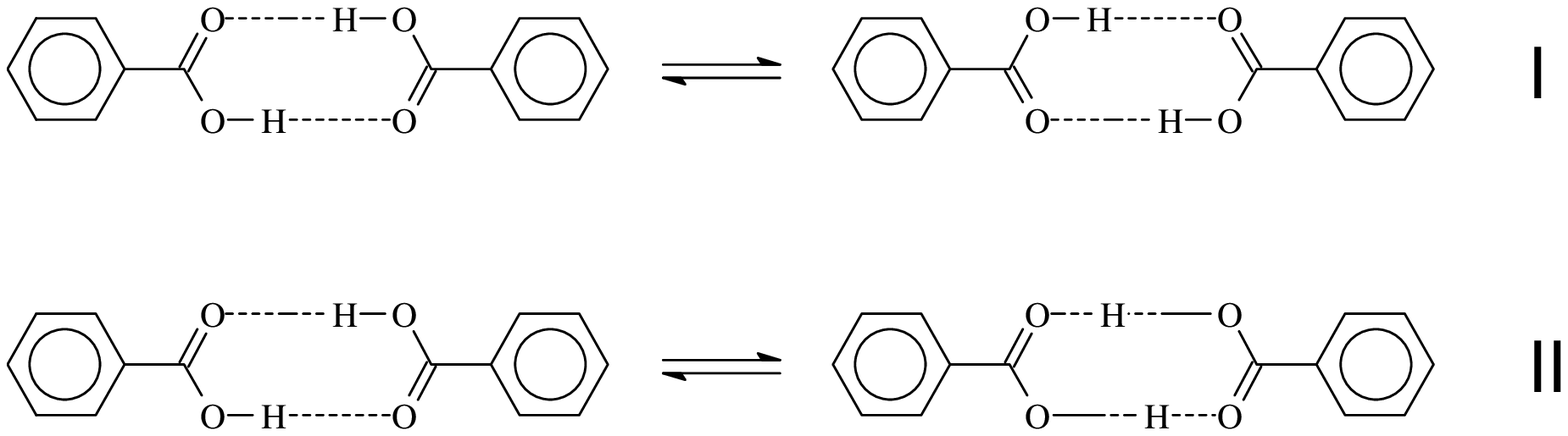}
\end{center}
\caption{\label{fig:1} Sketch of tautomerism (I) and proton interconversion (II).}
\end{figure}

Our rationale is based upon fundamental laws of quantum mechanics and experimental facts for BA. The vibrational spectra, like those of KHCO$_3$, are consistent with adiabatic separation and asymmetric double-wells for the OH stretch modes ($\nu$OH). \cite{FLR,FRLL,FTP} In addition, the spectra reveal a phenomenon that is not observed for KHCO$_3$: the thermal population of a state at $h\nu_l = (54 \pm 6)$ \cm\ is correlated with the interchange of single and double CO-bonds (tautomerism), occurring without any change of the proton states. \cite{FLR} This is consistent with markedly different timescales for tautomerism (Fig. \ref{fig:1}, I) and adiabatic interconversion of protons (Fig. \ref{fig:1}, II), hereunder referred to as ``interconversion'' for short. By analogy with the popular ``phonon-induced tautomerism'', \cite{ST} it was conjectured that $h\nu_l$ could be a phonon such that the ground state should correspond to the preferred tautomer at low temperatures and the excited states to a mixture of both tautomers sharing the same $\nu$OH states. \cite{FLR,Fil7} However, neutron diffraction studies by Wilson et al. \cite{WSF1} report a temperature law for the proton-site occupancy factor (H-SOF) that is at variance with both the proposed model and NMR/QENS data. For example, the estimated $R$-SOF at 20 K is $\approx 13\%$ instead of $\approx 0$ given by the other techniques. In order to establish our theory upon an unquestionable experimental basis, we have carried out new measurements with neutron diffraction of the crystal parameters at various temperatures.

In Sec. \ref{sec:2}, we report the structural parameters measured between 6 and 293 K and we show that there is no statistical disorder for protons. The reported H-SOF's are significantly different from those of Wilson et al. and more accurate. They are consistent with other observations. In addition, the quasi-linear correlation with the mean CO-bond lengths is a new information. Van t'Hoff laws for protons and CO-bonds compare favourably with spectroscopic data. The quantum theory is presented in Sec. \ref{sec:3}. We insist that vibrational spectra are representative of double-well operators for nonlocal observables, which are fundamentally different in nature from potential functions of classical coordinates. In \ref{sec:31} we set the Bloch wavefunctions for a lattice of bosons and we establish a new energy level scheme consistent with the new diffraction data. In \ref{sec:32} we show that antisymmetrization for a lattice of fermions yields a macroscopic-scale time-independent state with remarkable properties. In \ref{sec:341} we show that the crystal is immune to decoherence and we introduce the quantum-measurement scheme. Then, the theoretical temperature law for the $R$-SOF is compared with measurements in \ref{sec:34}. In \ref{sec:33} we show that NMR/QENS data are consistent with a coherent decay of measurement-induced transitory states.

\section{\label{sec:2}Neutron diffraction}

Data were collected with the four-circle diffractometer, 5C2, at the Orph\'{e}e reactor (Laboratoire L\'{e}on Brillouin). \cite{LLB} Needle-shaped single-crystals ($3\times 3 \times 10$ mm$^3$) were obtained by slow cooling of an ethanol-hexane solution. A specimen wrapped in aluminum was loaded in a closed-cycle-refrigerator whose temperature was controlled to within $\pm 1$ K. The structural parameters listed in Table \ref{tab:1} were obtained with CRYSTALS. \cite{CRYSTALS} Positional and thermal parameters and the SOF's ($\rho_L$, $\rho_R$) were allowed to vary independently.

\subsection{Crystal structure}

Inspection of intensities for absent reflections confirms the $P2_1/c$ ($C_{2h}^5$) space group assignment at every temperature. There is no evidence of disorder or domains. The unit cell parameters increase slightly with the temperature but there is no obvious discontinuity suggesting a phase transformation. There are four equivalent C$_6$H$_5$COOH entities per unit cell. Centrosymmetric dimers are linked through hydrogen bonds whose lengths 2.61 \AA\ $< R_\mathrm{OO} < 2.65$ \AA\ are slightly longer than those of KHCO$_3$ in the same temperature range ($2.58-2.60$ \AA). The mean distances between $R$- and $L$-sites are virtually temperature independent: ${L-L} = {R-R} \approx 2.24$ \AA\ and ${L-R} \approx 0.70$ \AA\ ($\approx 0.60$ \AA\ for KHCO$_3$). The aromatic-ring is practically temperature-independent, the CC-bonds are equivalent and the angles equal 120$^\circ$.

Wilson et al. have pointed out that dimers in planar layers are connected to each other by contacts between oxygens and hydrogens of the benzene ring. \cite{WSF2} These H$\cdots$O distances ($a$ and $b$ in Table \ref{tab:1}) increase slightly with temperature but there is no obvious correlation with SOF's or bond lengths.

\subsection{Quantum order versus classical disorder}

The structural model proposed by Wilson et al., is based on conflicting premises: Fractional SOF's are supposed to be representative of a statistical distribution of mutually exclusive empty or occupied sites and, at the same time, they are constrained by $\rho_L (T) + \rho_R(T) \equiv 1$. This is simply incorrect because a random distribution of occupied and empty sites should transfer some Bragg-peak intensity to a background of coherent diffuse scattering leading to an apparent decrease of the number of protons effectively detected. \cite{SWL,NK} At elevated temperatures, this would give $\rho_L \approx \rho_R \approx 0.25$, instead of 0.50 measured. In fact, our own refinements show that $\rho_L (T) + \rho_R(T)$ converges to 1 without any constraint and fractional SOF's can be rationalized without recourse to disorder with the partial differential nuclear cross-section in the quantum regime: \cite{SWL}

\begin{equation}\label{eq:1}
\frac{\mathrm{d}^2\sigma}{\mathrm{d}\Omega\mathrm{d}E} = \frac{k_f}{k_i} \sum\limits_{i,f} p_i |\langle\mathbf{k}_{nf}|\langle f|\hat{V}(\mathbf{r})|i\rangle|\mathbf{k}_{ni}\rangle|^2 \delta(\hbar\omega + E_i -E_f) .
\end{equation}
The summation runs over initial states $|i\rangle$, $E_i$, weight $p_i$, and final states $|f\rangle$, $E_f$. The neutron wave vectors are $\mathbf{k}_{ni}$ and $\mathbf{k}_{nf}$. The energy transfer is $\hbar \omega$ and $\mathrm{d}\Omega$ is an element of solid angle. The interaction operator, $\hat{V}(\mathbf{r})$, comprises operators at every site. For diffraction, $|i\rangle \equiv |f\rangle$, $E_i \equiv E_f$, $\hbar\omega \equiv 0$. Thanks to adiabatic separation, $\rho_L(T) + \rho_R(T) \equiv 1$ is representative of distinct sublattices of $LL$ or $RR$ dimers, respectively.

Alternatively, should the space periodicity be destroyed by disorder, the mean potential $\overline{\sum\limits_{i} p_i \langle i|\hat{V}(\mathbf{r})|i\rangle}$ could be Fourier transformed into the unit-cell structure factor and only the mean position of protons would be relevant. \cite{SWL} This model was graphically illustrated by Wilson et al. \cite{WSF2} who showed that protons located at $L$-sites at low temperature shift smoothly toward the center of the O$\cdots$O bonds as the temperature increases, while the anisotropy of the thermal ellipsoid increases tremendously along O$\cdots$O, as if the local potential were dramatically flattened. In contrast, the vibrational spectra do not show any significant frequency shift. This model is untenable and so is statistical disorder.

\subsection{\label{sec:2c}Temperature effects}

\begin{figure}[!hbtp]
\begin{center}
\includegraphics[scale=0.4, angle=0]{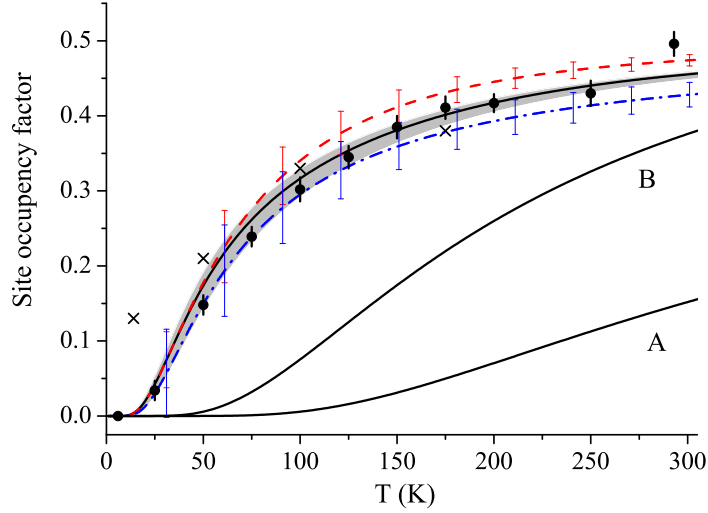}
\end{center}
\caption{\label{fig:2} (Color on line) Temperature laws for the occupancy factor of the less favored proton sites. $\bullet$ with error bars: experimental points (this work). $\times$: experimental points after Ref. [\onlinecite{WSF1}]. Blue dot dashed with bars: SOF due to tautomerism only. Solid plus grey zone: Tautomerism plus interconversion without the intermediate states (\ref{eq:13}). Red dashed with bars: Tautomerism plus interconversion including the intermediate states (\ref{eq:15}). The grey zone and the bars represent dispersions of the theoretical curves due to uncertainties for the parameters. Curves A and B are analogous to (\ref{eq:13}) and (\ref{eq:15}), respectively, for a single tautomer state (see text). }
\end{figure}

Fig. \ref{fig:2} shows substantial differences between $\rho_R(T)$ in Table \ref{tab:1} (solid circles) and the $R$-SOF's of Wilson et al. (X), \cite{WSF1,WSF2} especially at low temperatures. Our estimate $\rho_R(T) = 0$ at 6 K is in accordance with a well defined minimum of the potential energy. There is no more conflict with other techniques. The best fit to our data is consistent with a van t'Hoff equation substantially different from that of Wilson et al.:
\begin{equation}\label{eq:2}
\ln \frac{\rho_R} {\rho_L} = -0.07 \pm 0.03 + (86.8 \pm 1.7) T^{-1}.
\end{equation}
The difference in enthalpy $\Delta H_\rho = (86.8 \pm 1.7)$ K compares with the NMR-based estimate: $\Delta H_{NMR} = (86.5 \pm 1.5)$ K. \cite{NBMJHT}

\begin{figure}[!hbtp]
\begin{center}
\includegraphics[scale=0.4, angle=0]{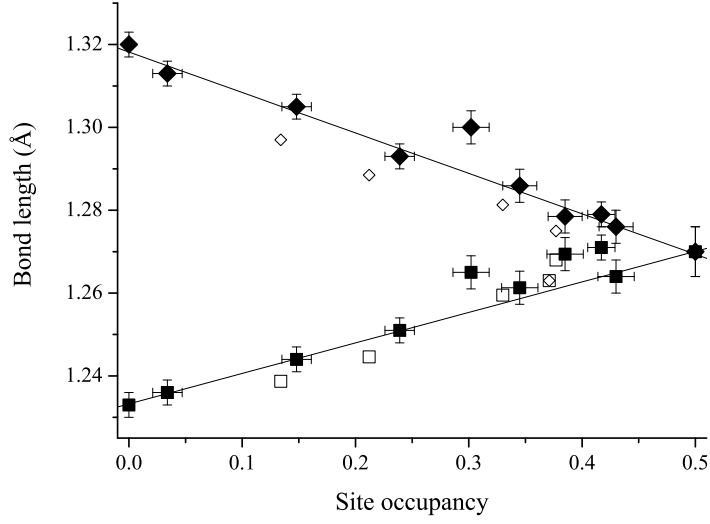}
\end{center}
\caption{\label{fig:3} CO bond lengths as a function of the occupancy of the less favored proton sites. $\blacksquare$ and $\blacklozenge$ with error bars: experimental points (this work). $\square$ and $\lozenge$: after Ref. [\onlinecite{WSF2}]. The straight lines are guides for the eyes.}
\end{figure}

Fig. \ref{fig:3} suggests a linear correlation of the CO bond lengths and $\rho_R$, apart from the points at 100 K ($\rho_R \approx 0.30$) for which the bonds are longer than expected by $\approx 0.01$ \AA, what is about twice the variance. Since $\rho_R$ at 100 K in Fig. \ref{fig:2} is not singular, there is no reason to question the data acquisition. Statistical errors could be actually greater than the variance given by CRYSTALS.

\begin{figure}[!hbtp]
\begin{center}
\includegraphics[scale=0.4, angle=0]{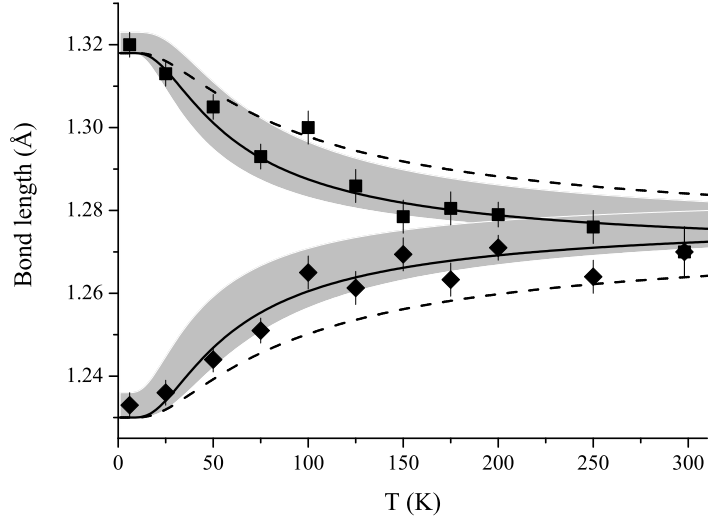}
\end{center}
\caption{\label{fig:4} CO bond lengths as a function of the temperature. $\blacksquare$ and $\blacklozenge$ with error bars: experimental points from Table \ref{tab:1}. Solid curves were calculated with (\ref{eq:4}) and $R_\mathrm{C=O} = 1.230(5)$ \AA, $R_\mathrm{C-OH} = 1.318(5)$ \AA, $\Delta E = (78 \pm 20)$ K$^{-1}$. The grey zones represent the statistical dispersion. The dashed curves (\ref{eq:5}) hold for the coexistence of both tautomers at elevated temperature [\onlinecite{Fil7}].}
\end{figure}

Because we know from spectroscopy that the frequencies of the COO modes are temperature-independent, \cite{FLR,FRLL} the increase of the C=O length and the decrease of the C-O bond do not correlate with a continuous change from the carboxylic groupings COOH to the carboxylate COO$^- $H$^+$. The bond lengths are more likely representative of an unresolved mixture of tautomer sublattices with constant CO-bond lengths. Then, with $R_{C=O} = 1.233$ \AA\ and $R_{C-OH} = 1.320$ \AA\ (Table \ref{tab:1}, 6 K), best fit exercises give:

\begin{equation}\label{eq:3}\begin{array}{c}
\ln \displaystyle{\frac{R_{C-OH} - R_{C1O1}(T)} {R_{C1O1}(T) -R_{C=O}}}  = 0.01 \pm 0.10 + (85.1 \pm 6.4)T^{-1};\\
\ln \displaystyle{\frac{R_{C=O} - R_{C1O2}(T)} {R_{C1O2}(T) -R_{C-OH}}}  = 0.18 \pm 0.15 + (71.7 \pm 9.5)T^{-1}.\\
\end{array}\end{equation}
As these equations should represent the same equilibrium, we take the mean values $\Delta S_{CO} = 0.1\pm 0.2$, $\Delta H_{CO} = (78 \pm 20)$ K, and the measured bond lengths are represented by a two-level system as:

\begin{equation}\label{eq:4}
\begin{array}{c}
R_\mathrm{C1O1} (T) = p_0 (T) R_\mathrm{C=O} + p_1 (T) R_\mathrm{C-OH}; \\
R_\mathrm{C1O2} (T) = p_0 (T) R_\mathrm{C-OH} + p_1 (T) R_\mathrm{C=O};
\end{array}
\end{equation}
with $p_0(T) = (1 + e^{-\Delta H_{CO}/T})^{-1}$ and $p_1(T) = p_0(T) e^{-\Delta H_{CO}/T}$. The solid lines in Fig. \ref{fig:4} and the dispersion zone (grey) representative of statistical errors compare reasonably with the experimental points. The mean deviation is less than $5\times10^{-3}$ \AA\ for each branch. Alternatively, a harmonic energy level scheme for a phonon such that the excited states be comprised of both tautomers in equal proportions, \cite{FLR,FRLL,Fil7} should give:
\begin{equation}\label{eq:5}
\begin{array}{rcl}
R_\mathrm{C1O1} (T) & = & p'_0 (T) R_\mathrm{C=O} + p'_1 (T) (R_\mathrm{C=O} + R_\mathrm{C-OH})/2; \\
R_\mathrm{C1O2} (T) & = & p'_0 (T) R_\mathrm{C-OH} + p'_1 (T) (R_\mathrm{C=O} + R_\mathrm{C-OH})/2;
\end{array}
\end{equation}
with $p'_0(T) = (1 - e^{-\Delta H_{CO}/T})^{-1}$ and $p'_1(T) = e^{-\Delta H_{CO}/T}$. The dashed curves in Fig. \ref{fig:4} is less satisfactory. In fact, it is unlikely that a single phonon could drive at once the many degrees of freedom involved in tautomerism. In addition, a flip-flop of the carboxylic group is sterically forbidden. Presumably, the monoclinic crystal field separates the ground states of the otherwise degenerate tautomer sublattices and $\Delta H_{CO}$ is not representative of an asymmetric double-well along a definite reaction path.

Finally, the various estimates $\Delta H_{CO}$, $\Delta H_{\rho}$, $\Delta H_{NMR}$, and $h\nu_l = (81 \pm 9)$ K, \cite{FLR,FRLL} yield the mean value of $\Delta H = (78 \pm 20)$ K. In Fig. \ref{fig:2}, the dot-dashed curve $\rho_R = p_1(T)$ is slightly underestimated above $\approx 150$ K, what suggests that adiabatic interconversion (Fig. \ref{fig:1}, II) is not negligible. The purpose of the quantum theory presented below is to calculate this contribution on a physically sound ground.

\section{\label{sec:3}Quantum theory}

The stretching modes play the leading role for adiabatic interconversion. In classical physics, a double-well between $L_1L_2$ and $R_1R_2$ is permitted for the symmetric coordinate $x_g$ (Raman-active), while it is excluded for the antisymmetric $x_u$ (infrared-active). The effective mass $\mu_g = 2$ amu is eventually increased by a promoter mode involving heavy nuclei. \cite{Luckhaus} However, knowing that classical trajectories are alien to quantum mechanics, vibrational spectra effectively reveal the eigenstates of observables, which are not classical coordinates. For a periodic lattice of indistinguishable centrosymmetric proton sites, the observables are symmetry species at the center of the Brillouin-zone (CBZ), say $\chi_{\xi}, \eta_\xi, \zeta_\xi,$ with $\xi =$ ``$u$'' or ``$g$'', for the stretching, in-plane bending, out-of-plane bending modes, respectively, and it is an experimental fact that both infrared and Raman spectra are consistent with double-well operators $\hat{V}_\xi(\chi_\xi)$, and single-well operators $\hat{V}_\xi(\eta_\xi)$, $\hat{V}_\xi(\zeta_\xi)$, for proton oscillators whose effective masses are $\mu_u = \mu_g = 1$ amu (see Table \ref{tab:2}). The INS spectrum is also consistent with this interpretation and reveal negligible dispersion for the OH modes. The eigenstates are, therefore, entirely determined.

\subsection{\label{sec:31}The OH states}

For the ease of notation, an ideal disorder-free crystal of BA can be represented by $\mathcal{N}$ reduced unit-cells indexed $j$ in 3-dimension, each of them comprising only one planar dimer. $\mathcal{N}$ is on the order of Avogadro's constant. With respect to the center of symmetry, the mean positional parameters of the bonding protons are $ \{x_{0,j}, y_{0,j}, 0\}$ for $L_{1,j}$, $\{-x_{0,j}, -y_{0,j}, 0\}$ for $L_{2,j}$, $\{-x_{0,j}, y_{0,j}, 0\}$ for $R_{1,j}$, $\{x_{0,j}, -y_{0,j}, 0\}$ for $R_{2,j}$. Small displacements are represented by local (classical) variables $s_{1,j},s_{2,j}$ $(s = x,y,z)$, and the mass- and distance-conserving normal coordinates are:
\begin{equation}\label{eq:6}\begin{array} {rclrcl}
s_{L_u,j} & = & \displaystyle{\frac{1}{\sqrt{2}}(s_{1,j} + s_{2,j}- 2s_{0,j})}; & s_{L_g,j} & = & \displaystyle{\frac{1}{\sqrt{2}}(s_{1,j} - s_{2,j})}; \\
s_{R_u,j} & = & \displaystyle{\frac{1}{\sqrt{2}}(s_{1,j} + s_{2,j}+ 2s_{0,j})}; & s_{R_g,j} & = & \displaystyle{\frac{1}{\sqrt{2}}(s_{1,j} - s_{2,j})}; \\
\end{array}\end{equation}

In classical physics, the effective oscillator mass of a normal mode is arbitrary. In quantum physics, normal coordinates are observables and their effective masses are definite and measurable. \cite{IKSYBF}

For each symmetry species, there are $\mathcal{N}$ Bloch states indexed $r$ ($1 \leq r \le \mathcal{N}$) whose wavefunctions are linear combinations of those of the unit-cells. For the lower states of the double-well $\hat{V}_\xi(\chi_\xi)$, they can be written as:
\begin{equation}\label{eq:7}
\begin{array}{rcl}
\psi_{0\xi r} (\mathbf{k}_{0r}, t, \varphi_{0r}) & = & \displaystyle{\frac {1} {\sqrt{\mathcal{N}}}}\sum\limits_{j=1}^\mathcal{N} [ \ \cos\theta_\xi\ \psi_{L\xi,j} + \sin\theta_\xi\ \psi_{R\xi,j} ] \exp i [\mathbf{k}_{0r} \mathbf{.H}_j + \omega_{0\xi r} t + \varphi_{0\xi r}] ;\\
\psi_{1\xi r} (\mathbf{k}_{1r }, t, \varphi_{1r}) & = & \displaystyle{\frac {1} {\sqrt{\mathcal{N}}}}\sum\limits_{j=1}^\mathcal{N} [ -\sin\theta_\xi\ \psi_{L\xi,j} + \cos\theta_\xi\ \psi_{R\xi,j} ] \exp i [\mathbf{k}_{1r} \mathbf{.H}_j + \omega_{1\xi r} t + \varphi_{1\xi r}] . \\
\end{array}
\end{equation}

$\mathbf{k}_{nr}$ is a wave vector; $\mathbf{H}_j$ is a vector of the sublattice; $\varphi_{n\xi r}$ is an arbitrary phase. \footnote{In contrast to previous works, \cite{FCKeen} the time variable is included explicitly for further examination} As there is no permanent electric dipole for centrosymmetric dimers, dispersion is negligible and the states are degenerate with respect to $\mathbf{k}_{nr}$. Each wavefunction spans every proton site and the occupation factor at a given site includes contributions from every state. $\psi_{L\xi,j}$ and $\psi_{R\xi,j}$ are harmonic eigenfunctions in 3-D for second-order expansions of the potential operator $\hat{V}_\xi(\chi_\xi) + \hat{V}_\xi(\eta_\xi) + \hat{V}_\xi(\zeta_\xi)$ around each minimum. Even at 300 K, only the ground states of the bending modes are significantly populated and the corresponding quantum numbers are omitted for the ease of notation. Because of the double-well, the symmetry species $\chi_\xi$ are not linear combinations of normal coordinates (\ref{eq:6}) and, conversely, (\ref{eq:7}) contains no information on the local coordinates $x_{1,j},x_{2,j}$.

The eigenstates of the $\nu$OH modes given in Table \ref{tab:2} are consistent with: $h\nu_{01_u} = h\nu_{01_g} = h\nu_{01}$, $\theta_u = \theta_g = \theta$, $\tan 2\theta = \nu_{t}/(\nu_{01} - \nu_{t})$, where $h\nu_{t} = (6 \pm 1)$ \cm\ is the tunnel splitting calculated numerically for the symmetric double-wells. Then, $\cos\theta \approx 1$ and $\sin\theta = \varepsilon = (1.8\pm 0.3)10^{-2}$. \cite{Fil7} The lack of $u-g$ splitting for $h\nu_{01}$, as opposed to higher transitions, is a consequence of the marked localization of the wavefunctions in either well. Creation(annihilation) of a quantum $h\nu_{01}$ can be, therefore, conceived of as the transfer of a particle-like entity from one minimum $L_\xi(R_\xi)$ to the other one $R_\xi(L_\xi)$, without any energy transfer to the internal degree of freedom.

\begin{figure}
\includegraphics[angle=0.,scale=0.55]{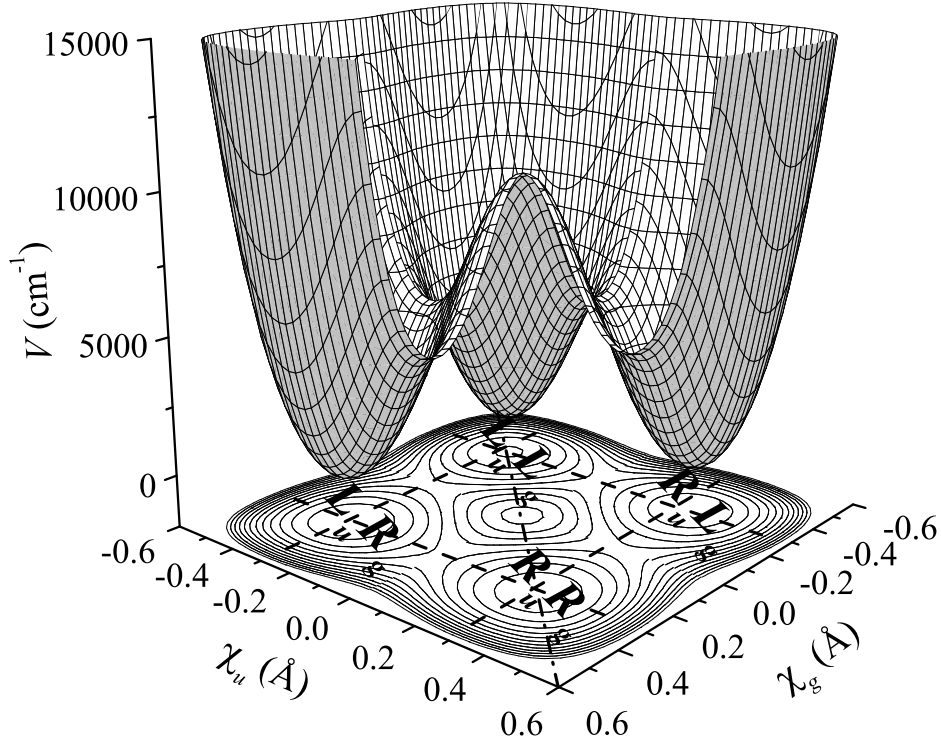}
\caption{\label{fig:5} Perspective view of the potential operator for the hydrogen bonding protons in the crystal of benzoic acid. $\chi_{u}$ and $\chi_{g}$ are the symmetry species of the OH stretching mode unveiled by vibrational spectra. }
\end{figure}

In classical physics, interconversion is thought of as an equilibrium between dimers $L_1L_2$ and $R_1R_2$, knowing that $L_1R_2$ and $R_1L_2$ are energetically excluded. Each dimer entity is supposed to be at any given time in a definite configuration, whether it is observed or not (classical realism). In quantum physics, symmetry species yield a much richer 4-level scheme. For the preferred tautomer sublattice at low temperatures (say $\mathfrak{T}_0$, labeled with the quantum number $l = 0$), the four-well operator $\hat{V}_0 (\chi_{u,j},\chi_{g,j}) = \hat{V}_{0u} (\chi_{u,j}) + \hat{V}_{0g} (\chi_{g,j})$ is shown in Fig. \ref{fig:5}. The central barrier of $\approx 10^4$ \cm\ is twice the barrier in 1-D. It is surrounded by four deep minima: $L_uL_g$ set to zero-energy; $R_uL_g$ and $L_uR_g$ at $\approx h\nu_{01}$; $R_uR_g,$ at $\approx 2h\nu_{01}$. Interconversion between $L_uL_g$ and $R_uR_g$ occurs through $R_uL_g$ and $L_uR_g$ at a very low energy cost compared to the classical mechanism, via $R_1L_2$, $L_1R_2$. For the less favored tautomer sublattice ($\mathfrak{T}_1$, $l = 1$), the C$-$O and C$=$O bonds are interchanged. The operator $\hat{V}_1 (\chi_{u,j},\chi_{g,j}) = \hat{V}_{1u} (\chi_{u,j}) + \hat{V}_{1g} (\chi_{g,j})$ is obtained by changing the signs of the linear terms, while the other terms are unaffected. Consequently: $R_uR_g$ is more stable that $L_uL_g$; (\ref{eq:7}) transforms as $\theta \longrightarrow \pi/2 -\theta$; $h\nu_{01}$, $h\nu_{02_\xi}$, $h\nu_{03_\xi}$, are unchanged.

\begin{figure}
\includegraphics[angle=0.,scale=0.55]{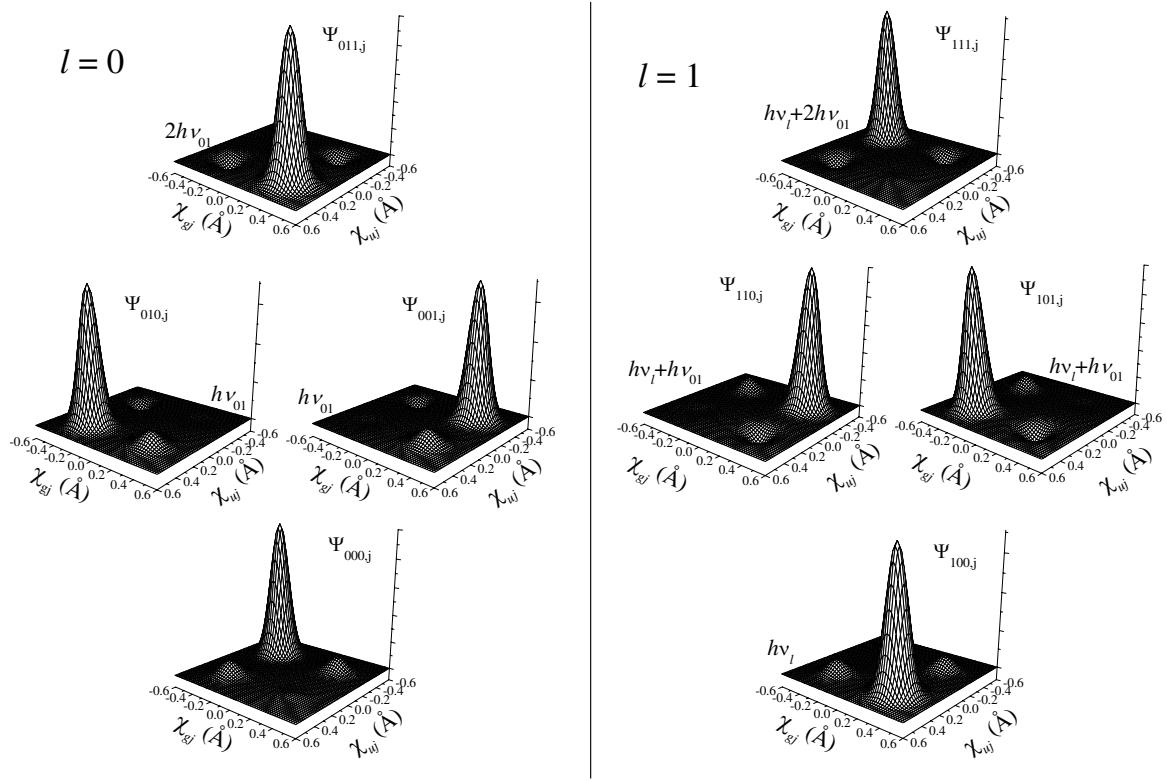}
\caption{\label{fig:6} Schematic view of the wave functions of the unit cell $j$ for the hydrogen bonding protons in the crystal of benzoic acid. $\chi_{uj}$ and $\chi_{gj}$ are the symmetry species of the unit cell unveiled by vibrational spectra. For the sake of clarity, the weak component of the wave function in one-dimension is multiplied by a factor of 5. The quantum numbers are $l,n_u,n_g$, respectively. }
\end{figure}

The scheme presented in Fig. \ref{fig:6} is consistent with the diffraction data reported in Sec. \ref{sec:2} (Fig. \ref{fig:4}). Compared to that previously published, \cite{FLR} the main difference is that the mixture of $\mathfrak{T}_0$ and $\mathfrak{T}_1$ for every $l \ge 1$ is now replaced by a two-level scheme: $l = 0$ for $\mathfrak{T}_0$ and $l = 1$ for $\mathfrak{T}_1$. The wavefunctions are:
\begin{equation}\label{eq:8}\begin{array}{l}
\mathcal{F}_{ln_un_g} (\mathbf{k}_{n_un_gr}, t, \varphi_{ln_un_gr}) = \displaystyle{\frac {1} {\sqrt{\mathcal{N}}}}\sum\limits_{j}^\mathcal{N}\Psi_{ln_un_g,j} \exp i [\mathbf{k}_{n_un_g r} \mathbf{.H}_j + \omega_{l n_un_g} t + \varphi_{ln_un_gr}];
\end{array}\end{equation}
where $\Psi_{ln_un_g,j} = \psi_{ln_u,j} \psi_{ln_g,j}$. \footnote{The energy level scheme of KHCO$_3$ ($h\nu_l \gg kT$) is similar to Fig. \ref{fig:6}-lhs with $h\nu_{01} = 216$ \cm\ and $\varepsilon \approx 0.05$.} The probability density $|\Psi_{000,j}|^2$, or $|\Psi_{011,j}|^2$, is localized to order $\varepsilon^2$ around $L_uL_g$, or $R_uR_g$, respectively. The mean geometry is $L_1L_2$ or $R_1R_2$. By contrast, $|\Psi_{010,j}|^2$ and $|\Psi_{001,j}|^2$ have no counterpart in classical physics. They represent counterintuitive $D_{2h}$ configurations with equal occupation factors at each of the four sites. This is markedly different from the $D_{2h}$ transition state of computational chemistry, with two protons at the center of the O$\cdots$O bonds. For $l = 1$, energy levels are shifted upwards by $h\nu_l$ and the wave functions are sketched in Fig. \ref{fig:6}-rhs.

\subsection{\label{sec:32}The crystal states}

Bloch states obey the Bose statistics law. If the fermion character of protons can be ignored, the sublattice can be represented by a superposition of degenerate states:
\begin{equation}\label{eq:9}
|\mathbb{B}\rangle = \sum\limits_{l,n_u,n_g} a_{ln_un_g} \sum\limits_{r} |\mathcal{F}_{ln_un_g} (\mathbf{k}_{n_un_gr}, t, \varphi_{ln_un_gr})\rangle;
\end{equation}
$a_{l00}^2 = p^l_l (1+p_l)^{-1} (1 + p_{01})^{-2}$; $a_{l10}^2 = a_{l01}^2 = a_{l00}^2 p_{01}$; $a_{l11}^2 = a_{l00}^2 p_{01}^2$; $p_l = \exp(-h\nu_l/kT)$; $p_{01} = \exp(-h\nu_{01}/kT)$. Then, destructive interferences cancel the phase factor, in space and time, so the probability density is zero everywhere, at any time, the Bloch states do not exist and the sublattice of protons is indefinite.

Alternatively, when the fermion character of bare protons  emerges from both the ionic character of the hydrogen bond and the adiabatic separation, Bloch states are forbidden by the exclusion principle. For a lattice of noninteracting bare protons (no spin-spin or electric-charge interaction, no exchange or Fermi energy) protons behave as sharply localized particles whose wavefunctions, say $\Phi (\mathbf{r}-\mathbf{r}_{0j})$ at a given site in a given unit cell, do not overlap: $\int \Phi (\mathbf{r}-\mathbf{r}_{0j'}) \Phi (\mathbf{r}-\mathbf{r}_{0j}) d\mathbf{r} = \delta_{jj'}$. This yields zero-width eigenfunctions, say $\mathcal{L}_{1j}, \mathcal{L}_{2j}, \mathcal{R}_{1j}, \mathcal{R}_{2j}$, for which mean-square amplitudes equal zero and tunneling is excluded. The sublattice of fermions is rigorously static and periodic (distortion-free). The centrosymmetric dimers $\mathcal{L}_{1j} \mathcal{L}_{2j}$ and $\mathcal{R}_{1j} \mathcal{R}_{2j}$ correspond to eigenstates $|00\rangle_j$ and $|11\rangle_j$, respectively, such that $E_{l11} - E_{l00} = (-1)^l 2h\nu_{01}$. The states $|\mathcal{R}_{1j} \mathcal{L}_{2j}\rangle_j$ and $|\mathcal{L}_{1j} \mathcal{R}_{2j}\rangle_j$ are excluded. There is no further eigenstate and the potential operator is irrelevant. In addition, the antisymmetry postulate for centrosymmetric pairs requires spin state-vectors with singlet-like antisymmetry, $|\mathcal{S}\rangle_j$, or triplet-like symmetry, $|\mathcal{T}\rangle_j$, \cite{CTDL} such as:
\begin{equation}\label{eq:10}\begin{array}{rcl}
|\mathbb{F} \rangle & = & \displaystyle{\frac{1}{\sqrt{2}}} \sum\limits_j \hat{S}_{21j} [(\alpha_{000} + \alpha_{111}) |00\rangle_j + (\alpha_{011} + \alpha_{100}) |11 \rangle_j ] |\mathcal{S}\rangle_j \\
& + & \displaystyle{\frac{1}{\sqrt{2}}} \sum\limits_j \hat{A}_{21j} [(\alpha_{000} + \alpha_{111}) |00\rangle_j + (\alpha_{011} + \alpha_{100}) |11\rangle_j] |\mathcal{T}\rangle_j;\\
\end{array}\end{equation}
$\hat{S}_{21j} = (1 + \hat{P}_{21j})/2$ is the symmetrizer; $\hat{A}_{21j} = (1 - \hat{P}_{21j})/2$ is the antisymmetrizer; $\hat{P}_{21j}$ is the permuter; \footnote{The degenerate spin-symmetry is totally different from the splitting of singlet and triplet states due to spin-spin interaction in magnetic systems} $\alpha_{l00}^2 = p_l^l (1 + p_l)^{-1} (1 + p_{01}^2)^{-1}$; $\alpha_{l11}^2 = \alpha_{l0}^2 p_{01}^2$. The thermal statistics are, therefore, different for fermions and bosons. However, energy transfer to the static lattice and interconversion between $\mathcal{L}_{1j} \mathcal{L}_{2j}$ and $\mathcal{R}_{1j} \mathcal{R}_{2j}$ are forbidden and it is necessary to suppose a thermally induced mechanism to account for thermal equilibration.

\subsection{\label{sec:341}Quantum measurements and decoherence}

In the first step of a quantum measurement, the system to be measured is entangled with another quantum object so as to realize, or prepare, a state that will be effectively measured. At the microscopic level of a single object, the back-action of the measurement yields unavoidable changes of the system due to the collapse of the wavefunction. In classical physics, by contrast, a measurement can be carried out with unlimited precision without affecting the system being measured (noninvasive measurement). From the viewpoint of measurements, a crystal of BA is both quantum and classical. First, an incoming wave can realize a pure state ($r$). Second, the pre-existing state is practically unaffected by the realization of a state amongst a quasi-infinite number of equally probable realizations. The back-action can be ignored. In addition, the remote detection of the outgoing wave has no back-action to the measured system. Third, the probability for an incident wave to interact with a previously realized state is insignificant. The evolution of the prepared state is determined by the time-dependent Schr\"{o}dinger equation, irrespective of further measurements. Fourth, both the pre-existing and the realized states depend on the temperature that is a macroscopic-scale classical variable of the crystal.

From the viewpoint of decoherence, the crystal is an open system in a gaseous atmosphere. Under normal conditions, the density of the surroundings is about 1000-fold less than that of the crystal, so the probability for an incoming thermal photon to realize a proton state is significant. However, decoherence is hampered because quantum correlations are constrained by the crystal structure. For bosons (\ref{eq:9}), creation/anihilation of a Bloch state does not induce decoherence. For fermions (\ref{eq:10}), there is noway to transfer any energy to the static state, so as to reach thermal equilibrium, other than via realization of a driven Bloch state violating the quantum laws for fermions. We show below that the lifetime of such states is short enough to avoid irreversible decoherence of the bulk.

In order to confront the theory with observations, we examine which experiments yield different outcomes for fermions or bosons. In the first instance, we can discard vibrational spectroscopy because the same Bloch states are realized via resonant energy transfer in both cases. By contrast, neutron diffraction and relaxometry provide experimental support to fermions.

\subsection{\label{sec:34}Neutron diffraction}

Elastic momentum transfer ($\Q = \mathbf{k}_{ni} - \mathbf{k}_{nf}$) realizes a Bloch state whose wave vector is $\Q$. For uncorrelated scattering events, the state spans the $\mathcal{N}$ unit cells, irrespective of the coherence length of the beam. In the boson case, the realized state is
\begin{equation}\label{eq:14}
|B_1\rangle_r = \sum\limits_{l,n_u,n_g} a_{ln_un_g} |\mathcal{F}_{ln_un_g} (\mathbf{Q},t)\rangle.
\end{equation}
Diffraction gives Bragg-peaks whose intensities are proportional to the nuclear coherent cross-section for protons ($\sigma_c \approx 1.8$ b) and to the Debye-Waller factor, an exponentially decreasing function of $\Q$ due to distortions. The occupancy factors are $L$-SOF $= a_{000}^2 +(a_{010}^2 + a_{001}^2 + a_{110}^2 + a_{101}^2)/2 + a_{111}^2 $, $R$-SOF $= a_{100}^2 +(a_{110}^2 + a_{101}^2 + a_{010}^2 + a_{001}^2)/2 + a_{011}^2 $, and $L$-SOF + $R$-SOF $\equiv 1$ at every temperature. The temperature law for $R$-SOF
\begin{equation}\label{eq:15}
\rho_{RB}(T) = \displaystyle{\frac{p_l + p_{01}} {(1 + p_l)(1 + p_{01})}}, \\
\end{equation}
gives the dashed curve in Fig. \ref{fig:2} that is overestimated compared to measurements.

For a lattice of fermions, the static Bloch state effectively realized depends on $\Q$. For $\Q = \mathbf{L}^* \ne \mathbf{H}^*$ ($\mathbf{L}^*$ is the reciprocal crystal lattice), $\Q$-induced distortions destroy the spin-symmetry:
\begin{equation}\label{eq:12}\begin{array}{rcl}
|H_1^+\rangle_r & = & \displaystyle{\frac{1}{\sqrt{\mathcal{N}}}}\sum\limits_j [(\alpha_{000} + \alpha_{111}) |00\rangle_{j} + (\alpha_{011} + \alpha_{100}) |11\rangle_{j}]\exp(i\mathbf{Q.L}_j). \\
\end{array}\end{equation}
The diffraction pattern is basically the same as that anticipated for bosons. However, the occupancy factors are different: $L$-SOF $= \alpha_{000}^2 + \alpha_{111}^2$; $R$-SOF $= \alpha_{100}^2 + \alpha_{011}^2$; $L$-SOF + $R$-SOF $\equiv 1$. The temperature law for $R$-SOF,
\begin{equation}\label{eq:13}
\rho_{RF} (T) = \frac {p_l + p_{01}^2} {(1 + p_l) ( 1 + p_{01}^2) },
\end{equation}
gives the solid curve and the grey dispersion zone (Fig. \ref{fig:2}) in better agreement with experiments than $\rho_{RB} (T)$. \footnote{For the point at 293 K, it is not clear whether there is a real discontinuity above 250 K or whether statistical errors due to the limited number of useful reflections were underestimated}

For the sake of comparison with KHCO$_3$, the curves A and B in Fig. \ref{fig:2} represent (\ref{eq:13}) and (\ref{eq:15}), respectively, for a single tautomer ($h\nu_l \gg kT$). These curves are markedly different and, for KHCO$_3$, measurements are unambiguously in accordance with fermions (\ref{eq:13}), whereas bosons (\ref{eq:15}) are excluded. \cite{FCG2,Fil7} For BA, the difference is not so dramatic, because of the tautomer statistics, but the same trend in favour of fermions is observed.

It should be mentioned that for $\mathbf{Q} = \mathbf{H}^*$ the Bloch state $|H_2^+\rangle_r = |\mathbb{F}\rangle/\sqrt{\mathcal{N}}$ retains the essential features of the pre-existing state: static periodicity and spin-symmetry. Then, the cross-section for coherent scattering is the total cross-section including the huge incoherent scattering cross-section ($\sigma_i \approx 80$ b): $\sigma = \sigma_i + \sigma_c \approx 82.0$ b. \cite{SWL} The enhanced diffraction pattern $\mathbf{H}^*$ superimposed to the regular pattern $\mathbf{L}^*$ is an evidence of the spin-symmetry featuring fermions. In addition, because the Debye-Waller factor equals $1$ at every temperature, the contrast is best at large $\Q$-values and at elevated temperatures. This was effectively observed for KHCO$_3$ and, as expected, not for KDCO$_3$. So far, there is no such experiment ever reported for BA.

\subsection{\label{sec:33}Relaxometry}

In classical physics, relaxometry techniques are thought of as noninvasive measurements of predetermined fluctuations. \cite{MGE,NBMJHT,SWL} QENS is specific to protons, for incoherent scattering by other nuclei is negligible. This is not so stark for NMR because the spin-lattice relaxation rate ($T_1^{-1}$) depends on many interactions. Nevertheless, both techniques give for BA comparable relaxation rates, which are essentially due to the bonding protons. \cite{NBMJHT} It is commonly supposed that uncorrelated jumps of proton pairs between $L_1L_2$ and $R_1R_2$ can be represented by a rate equation for the probability $P_L(t)$ that, given a $L_1L_2$-dimer at $t = 0$, the same dimer is $R_1R_2$ at time $t$, and conversely for $P_R(t)$. The stochastic process occurring over a relatively long timescale gives rise to a very weak Lorentzian profile in $\omega$, centered at $\omega = 0$, whose width is proportional to the inverse relaxation time $\tau^{-1}_H$. \cite{Springer,EGS,Bee,Hempelmann} This scheme is clearly in conflict with the rule stipulating that the interaction of a wave and a crystal is nonlocal.

In quantum physics, an incoming photon or neutron can realize a Bloch state of the stretching mode, whose wave vector is the momentum transfer $Q_x$ along the hydrogen bond ($Q_x \approx 0$ for NMR), while the bending modes can be ignored. For a lattice of bosons, the realized state analogous to (\ref{eq:14})
\begin{equation}\label{eq:16}\begin{array}{l}
|B_2^+\rangle_{r} = \sum\limits_{l,n_u,n_g} a_{ln_un_g}|\mathcal{F}_{ln_un_g}(Q_x,t)\rangle
\end{array}\end{equation}
is stationary and does not contribute to any relaxation mechanism, thanks to adiabatic separation. The rate effectively observed is a prima facie evidence that protons do not behave like bosons. For fermions, off-resonance energy transfer ($\hbar\omega \ll h\nu_l, h\nu_{01} \dots$) realizes a time-dependent state analogous to (\ref{eq:16}) and sets the starting time $t = 0$ for subsequent decay. In order to show that this decay can account for the observed rate, we consider, in the first instance, a simple superposition, say $a_{000}|000\rangle + a_{010}|010\rangle$ (from now on, we ignore $r$ and $|ln_un_g\rangle \equiv |\mathcal{F}_{ln_un_g}(Q_x,t)\rangle$). The probability density in the stationary regime is to order $\varepsilon^2$:
\begin{equation}\label{eq:17}\begin{array}{l}
|a_{000}|000\rangle + a_{010}|010\rangle|^2 \approx \\
a_{000}^2\psi_{u00_u} ^{2} \psi_{g00_g} ^{2} + a_{010}^2 \psi_{u01_u} ^{2} \psi_{g00_g} ^{2} + a_{000}a_{010}\varepsilon (\psi_{R_u}^2 - \psi_{L_u}^2) \psi_{L_g}^2 \cos2\pi\nu_{01}t. \\
\end{array}\end{equation}
Harmonic oscillations at frequency $\nu_{01}$ of $\varepsilon (\psi_{R_u}^2 - \psi_{L_u}^2)$ from $\varepsilon\psi_{R_u}^2/2$ to $-\varepsilon\psi_{R_u}^2/2$ in one well and from $-\varepsilon\psi_{L_u}^2/2$ to $\varepsilon\psi_{L_u}^2/2$ in the other well, during a half period $(2\nu_{01})^{-1}$, determine the inverse residence time: $2\varepsilon \nu_{01} = 2\nu_{t}$, in proton per second units ($Hs^{-1}$). (Similar terms are given in Table \ref{tab:3} for every binary combination.) Then, we define the relaxation rate as the critical decay of the normalized amplitude: $\tau_{0u}^{-1} = 2\varepsilon \nu_{01} {a_{010}}/ {a_{000}} = 2\varepsilon \nu_{01} p_{01}^{1/2}$, and likewise for $\tau_{0g}^{-1}$. The fractional exponent is distinctive of a coherent process.

Next, we consider the probability density $|\sum_{n_un_g} a_{n_un_g}|n_un_g\rangle|^2$ for a single tautomer $(p_l = 0)$. Because $Q_x$ cannot distinguish $u$ and $g$ species, we obtain two rates (to order $\varepsilon^2$): $|a_{00}|00\rangle + a_{01}|01\rangle + a_{10}|10\rangle|^2 $ gives $\tau_{0}^{-1} = \tau_{0g}^{-1} + \tau_{0u}^{-1} = 4\varepsilon\nu_{01}p_{01}^{1/2}$ and $|a_{01}|01\rangle + a_{10}|10\rangle + a_{11}|11\rangle|^2 $ gives
 \begin{equation}\label{eq:18}
\tau_{1}^{-1} = \tau_{1g}^{-1} + \tau_{1u}^{-1} = 4 \varepsilon \nu_{01} \displaystyle{\frac{a_{01}a_{11}}{a_{00}^2}}  = 4\varepsilon \nu_{01} p_{01}^{3/2}.
\end{equation}
In fact, because the four states are correlated, the coherent decay limited by the longer residence time $\tau_{1} > \tau_{0}$ gives the overall rate $2\tau_{1}^{-1} =8\varepsilon p_{01}^{3/2}$. This accords with the temperature law effectively measured for KHCO$_3$. \cite{EGS,Fil7}

\begin{figure}
\includegraphics[angle=0.,scale=0.4]{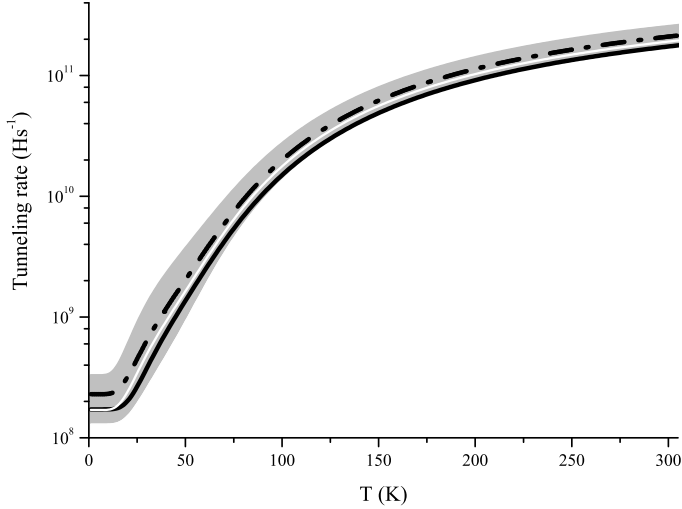}
\caption{\label{fig:7} Comparison of the experimental temperature law of BA (solid line) (\ref{eq:19}) with the quantum theory (dot-dashed) (\ref{eq:20}): $\nu_{01} = 5.16 \times 10^{12} s^{-1}$, $\varepsilon = 0.018$, $\tau_{Q}^{-1} = (2.3\pm1)10^8 \coth [(39\pm10)/T] + (6.77\pm1.15)10^{9}\{\exp[-(89\pm13)/T] + \exp[-(167\pm13)/T]\} + (7.2\pm1.2)10^{11} \exp[-(385\pm9)/T]$, in proton per second units. The grey zone features the dispersion of the theoretical curve due to uncertainties. The white curve was obtained with $\varepsilon = 0.016$. }
\end{figure}

For BA, NMR and QENS are both consistent with a markedly different empirical law: \cite{NBMJHT}
\begin{equation}\label{eq:19}
\begin{array}{rcl}
\tau_{exp}^{-1} & = & (1.72 \pm 0.02) 10^8 \coth[(43\pm 1)/kT] \\
& + & 10^{10} \exp(-120/kT) + 6.3\times 10^{11} \exp(-400/kT).
\end{array}\end{equation}
The first term was ascribed to the transition rate of the thermal bath supposed to rock the double-well asymmetry of $(86\pm 2)$ K for $LL$ and $RR$. The other terms were attributed to thermal jumps in the classical regime. Alternatively, Fig. \ref{fig:7} shows that (\ref{eq:19}) (solid curve) is consistent with the theoretical law for coherent decay of a Bloch state (dashed curve and grey zone):
\begin{equation}\label{eq:20}\begin{array}{rcl}
\tau^{-1}_{Q} & = & 8 \varepsilon^3 \nu_{01} \coth(h\nu_{l}/2kT) + 4 \varepsilon^2 \nu_{01} p_{01}^{1/2} \left[p_l^{-1/2} +  p_l^{1/2} \right] + 8 \varepsilon \nu_{01} p_{01}^{3/2} .\\
\end{array}\end{equation}
This function is different from that derived from the previous energy level scheme. \cite{Fil7} It contains more terms than (\ref{eq:19}) but it is more economic, since six empirical parameters are replaced by three scaling factors ($\nu_l, \nu_{01}, \varepsilon$) bridging relaxometry and spectroscopy.

The first theoretical term is formally identical to that in (\ref{eq:19}) but it is fundamentally different in nature. The rationale is that thermal transitions at $\pm h\nu_l$ occurring on a much shorter timescale than the interchange of single and double CO-bonds do not toggle the double-well asymmetry. The cross-section for one-quantum annihilation is the Bose factor $n(\nu_{l}) = [\exp(h\nu_{l}/kT) -1]^{-1}$, the cross-section for one-quantum creation is $[1+n(\nu_{l})]$ and the population ratio is $n(\nu_{l})/ [1+n(\nu_{l})] = p_l$. \cite{SWL} Annihilation yields $n(\nu_l)$ tautomer lattices $\mathfrak{T}_{1}$ in the lower energy level ($l = 0$) and creation yields $[1+n(\nu_l)]$ $\mathfrak{T}_{0}$ in the upper level ($l = 1$). The total population of $\mathfrak{T}_{0}$ and $\mathfrak{T}_{1}$ occupying the same level is $[1 + 2n(\nu_{l})] = \coth(h\nu_{l}/2kT)$, that is the mean energy in $h\nu_l$ units of the two-level system. Then, an incoming wave can realize
\begin{equation}\label{eq:21}\begin{array}{rcl}
|H_3^+\rangle = \coth(h\nu_{l}/2kT) |B_2^+\rangle.
\end{array}\end{equation}

The superposition of proton eigenstates for $\mathfrak{T}_{0}$ and $\mathfrak{T}_{1}$ occupying the same level,
\begin{equation}\label{eq:22}
\coth(h\nu_{l}/2kT) \sum_{n_un_g} a_{n_un_g}[|0n_un_g\rangle + |1n_un_g\rangle],
\end{equation}
gives rise to a tunnel splitting proportional to the overlap integral, $h\nu_\pm = h\nu_t \langle 0n_un_g | 1n_un_g\rangle = 4 \varepsilon^3 \nu_{01}$, and the decay rate is $2h\nu_\pm\coth(h\nu_{l}/2kT)$. We are not aware of any published example of a tunnel splitting promoted by thermal fluctuations, or, at very low temperatures, by quantum fluctuations of the zero-point energy. It is rather counterintuitive that a coherent superposition of proton states could emerge from an incoherent mixture of tautomer states. This is possible for fermions because the pre-existing state is insulated from environmental fluctuations and the coherent superposition of the four-well operators for $\mathfrak{T}_{0}$ and $\mathfrak{T}_{1}$ is realized at the time of the measurement for a seemingly static mixture of tautomers in the same state. For bosons, the same fluctuations would yield a mixture of the pre-existing four-well operators and no tunneling.

The second term in square brackets corresponds to combinations such as $a_{000} |000\rangle + a_{110} |110\rangle + a_{101} |101\rangle$ and $a_{100} |000\rangle + a_{010} |010\rangle + a_{001} |001\rangle$, which should not occur for bosons. The rate is independent of $\coth(h\nu_{l}/2kT)$ because it does not matter which $\mathfrak{T}_l$ is occupying the lower/upper level. The longest residence time is $(2\varepsilon^2 \nu_{01})^{-1}$ (see Table \ref{tab:3}) and the rate is proportional to $a_{110}/a_{000} = a_{101}/a_{000} = p_{01}^{1/2}p_l^{1/2}$ and $a_{010}/a_{100} = a_{001}/a_{100} = p_{01}^{1/2}p_l^{-1/2}$. The composite nature of this term is overlooked in (\ref{eq:19}) but Fig. \ref{fig:7} shows that the theoretical curve is close to observations in the intermediate temperature range ($50-150$ K). Finally, the last term of (\ref{eq:20}) is identical to (\ref{eq:18}).

The coherent decay of the real part of the wavefunction (\ref{eq:21}) accounts for the observed rate at every temperature. (The white curve in Fig. \ref{fig:7} shows that the agreement is improved with $\varepsilon = 1.6\times 10^{-2}$.) This deterministic evolution is not related to ``proton transfer'' in any way. The scaling factors $p_l^{1/2}, p_{01}^{1/2}, \varepsilon$, are effectively measured and numerics compare favorably with spectroscopic data. If this agreement is not fortuitous, environmental decoherence, transition rate induced by a thermal bath and semiclassical jumps should be excluded. In addition, the imaginary part of the wavefunction is fully determined by the measurement setting $Q_x$ and $t = 0$. We can thus obtain on a purely experimental basis the maximal knowledge that is theoretically available about the state. The lifetime of the Bloch state ($\sim 10^{-7} - 10^{-10}s$) is much longer than the period $\nu_{01}^{-1} = 2\times 10^{-13}s$ of the $\nu$OH mode and quasi-stationary quantum interferences occur on a macroscopic scale at every temperature.

Interesting enough, the splitting $h\nu_\pm \approx 8\times 10^{-3}$ \cm\ estimated for the crystal of BA compares favorably with those reported for isolated dimers in supersonic jet expansions for which $h\nu_l = 0$: $\approx 3\times 10^{-3} -16 \times 10^{-3}$ \cm\ for (DCOOH)$_2$; $\approx 12 \times 10^{-3}$ \cm\ for (HCCOH)$_2$; \cite{MH,*OH,*GSH} $\approx 8\times 10^{-3}$ \cm\ for formic-acetic acid (HCOOH...HOOCCH$_3$); \cite{TOH} $\approx 10^{-2}$ \cm\ for formic-propiolic  acid (HCOOH...HOOCCCH).\cite{DDSN} In these cases, the splitting should exist whether it is observed or not. The four-well operator of BA is apparently a feature of carboxylic-acid dimers, centrosymmetric or not, irrespective of the physical state. It is also rather counterintuitive that isolated dimers at very low temperature compare with BA in the high temperature regime: $kT\gg h\nu_l$.

\section{Conclusion}

The structure of the crystal of benzoic acid is consistent with a bistable system of unresolved tautomer sublattices sharing the same Hydrogen-bonding proton-sites and there is no evidence of disorder for these protons. Diffraction, vibrational spectroscopy and relaxometry techniques, provide evidences of wave-like protons excluding definite particles. The notions of reaction path across a potential function, transition state, stepwise or concerted proton transfer, are irrelevant.

We propose a quantum theory for the bonding protons regarded as fermions. The exclusion principle and the antisymmetry postulate yield a superposition of static lattice-states with spin-symmetry. According to the theory of quantum measurements, an incoming wave is supposed to realize a pure state without any significant back-action to the lattice state. For vibrational spectroscopy, resonant energy transfer to the pre-existing static lattice realizes the time-dependent eigenstates of a four-well operator in the abstract space of the symmetry species. For NMR and QENS, off-resonance energy transfer realizes a transitory state and the coherent decay rate of the real part of the wavefunction is measured. For diffraction, elastic momentum transfer realizes a static proton state in accordance with the site occupancies effectively measured. The theory accounts for every measurement with only three temperature-independent factors. Observations supporting a pre-existing lattice of fermion are the proton-site occupation factors revealed by diffraction, on the one hand, and, on the other, the spontaneous decay of Bloch states and the tunnel splitting promoted by environmental fluctuations. The alternative possibility that protons could behave like bosons when they are not observed can be ruled out.

Both quantum and classical physics hold for the nuclei of an open crystal of BA at every temperature. Macroscopic-scale quantum correlations rooted in the periodic structure cannot be destroyed and there is no dividing line between the quantum and the classical regime.

\clearpage

\addtolength{\hoffset}{-1.5cm}

\begin{turnpage}
\begin{table} \caption{\label{tab:1} Neutron single crystal diffraction data and structure refinement. $\lambda$ = 0.8305 \AA. Space group $P 2_1/c$. $\sigma$(I) limit: 3.00. Refinement on F. The occupation numbers for the proton sites are $\rho_L$ and $\rho_R$. The variance for the last digit is given in parentheses.}

\begin{tabular}{llllllllllll}
\hline
 Crystal data  & 6 K & 25 K & 50 K & 75 K & 100 K & 125 K & 150 K & 175 K & 200 K & 250 K & 293 K\\
\hline
$a$(\AA) & 5.401(1) & 5.401(1) & 5.415(1) & 5.415(1) & 5.500(1) & 5.427(1) & 5.456(1) & 5.456(1) & 5.500(1) & 5.500(1) & 5.500(1) \\
$b$(\AA) & 5.004(1)  & 5.004(1) & 5.016(1) & 5.016(1) & 5.100(1) & 5.112(1)& 5.074(1) & 5.074(1) & 5.100(1) & 5.100(1) & 5.100(1) \\
$c$(\AA) & 21.879(1)  & 21.879(1) & 21.826(1) & 21.826(1) & 22.020(1) & 21.810(1) & 21.875(1) & 21.875(1) & 22.020(1) & 22.020(1) & 22.020(1) \\
$\beta$  & 98.47(1)$^\circ$ & 98.47(1)$^\circ$ & 98.44(1)$^\circ$ & 98.44(1)$^\circ$ & 97.90(1)$^\circ$ & 98.26(1)$^\circ$ & 98.12(1)$^\circ$ & 98.12(1)$^\circ$ & 97.90(1)$^\circ$ & 97.90(1)$^\circ$ & 97.90(1)$^\circ$ \\
Volume (\AA$^3$)  & 584.8(2)  & 584.8(2) & 586.4(2) & 586.4(2) & 611.8(2) & 598.8(2) & 599.4(2) & 599.4(2) & 611.8(2) & 611.8(2) & 611.8(2) \\
Reflections measured & 2950  & 3594 & 2797 & 3364 & 3067 & 3092 & 2778 & 3607 & 3502 & 3828 & 3401 \\
Independent reflections & 2589 & 2637 & 1946 & 2645 & 2724 & 2283 & 2297 & 2300 & 2714 & 2748 & 2630 \\
Reflections used      & 1649 & 1679 & 1382 & 1472 & 1363 & 1122 & 1018 & 951 & 1031 & 755 & 511 \\
R-factor & 0.093 & 0.079 & 0.058 & 0.090 & 0.051 & 0.048 & 0.047 & 0.041 & 0.069 & 0.049 & 0.50 \\
Weighted R-factor & 0.114 & 0.097 & 0.068 & 0.051 & 0.053 & 0.048 & 0.049 & 0.037 & 0.045 & 0.032 & 0.030 \\
Number of parameters & 132 & 137 & 137 & 137 & 137 & 137 & 137 & 137 & 137 & 137 & 146 \\
Goodness of fit & 1.148 & 1.117 & 1.113 & 1.026 & 1.085 & 1.124 & 1.134 & 1.101 & 1.033 & 1.113 & 1.101 \\
Occupation $\rho_L$ & 1.00 & 0.966(13) & 0.852(13) & 0.761(13) & 0.69(1) & 0.65(1) & 0.61(1) & 0.59(1) & 0.583(12) & 0.57(2) & 0.50(1) \\
Occupation $\rho_R$ & 0.00 & 0.034(13) & 0.148(13) & 0.239(13) & 0.31(1) & 0.34(1) & 0.38(1) & 0.41(1) & 0.417(12) & 0.43(2) & 0.50(1) \\
$d_{LR}$ (\AA) & $--$ & $--$ & 0.68(1) & 0.71(1) & 0.69(1) & 0.69(1) & 0.67(1) & 0.72(1) & 0.70(1) & 0.69(2) & 0.68(1) \\
$R_\mathrm{OO}$ (\AA) & 2.623(5) & 2.621(5) & 2.619(5) & 2.614(5) & 2.646 (5) & 2.628(5) & 2.621(5) & 2.610 (5) & 2.624(5) & 2.623(5) & 2.629(5) \\
Bond length C1O1 (\AA) & 1.233(3) & 1.236(3) & 1.244(3) & 1.251(3) & 1.268(2) & 1.261(3) & 1.269(3) & 1.263(2) & 1.271(3) & 1.264(3) & 1.260(6) \\
Bond length C1O2 (\AA) & 1.320(3) & 1.313(3) & 1.305(3) & 1.293(3) & 1.302(1) & 1.286(3) & 1.279(3) & 1.280(3) & 1.279(3) & 1.276(4) & 1.266(6) \\
contact $a$ (\AA) & 2.546(3) & 2.545(3) & 2.531(3) & 2.530(3) & 2.549(4) & 2.522(3) & 2.536(3) & 2.541(4) & 2.565(3) & 2.563(5) & 2.610(6) \\
contact $b$ (\AA) & 2.462(3) & 2.466(3) & 2.480(3) & 2.490(3) & 2.536(4) & 2.517(3) & 2.528(3) & 2.534(4) & 2.548(3) & 2.553(8) & 2.553(6) \\
\hline
\end{tabular}
\end{table}
\end{turnpage}

\clearpage

\addtolength{\hoffset}{1.5cm}

\begin{table}
\caption{\label{tab:2} Double-well operators, distances between minima ($\Delta \chi_\xi$), barrier heights ($H_\xi$), and energy levels, for the crystal of the ring deuterated Benzoic acid. $\chi_\xi$ are the symmetry species infrared- ($\chi_u$) or Raman-active ($\chi_g$). The oscillator mass consistent with $\Delta \chi_\xi$ is 1 amu. *: observed with inelastic neutron scattering after Ref. [\onlinecite{PFJHT}]. }
\begin{tabular}{llcllll}
\hline
 & $\hat{V}_\xi(\chi_\xi)$ (\cm) & $\Delta \chi_0$ (\AA) & $H_\xi$ (\cm) & $h\nu_{01}^\xi$ (\cm) & $h\nu_{02}^\xi$ (\cm) & $h\nu_{03}^\xi$ (\cm) \\
\hline
Infrared \cite{FLR} & $265 \chi_u + 0.2860\times 10^6 \chi_u^2 $ & 0.70 & 5005 & $(172 \pm 4)$* & $2570 \pm 20$ & $2840 \pm 20$ \\
 & $+ 171480\exp(-2.17 \chi_u^2)$ & & & & & \\
Raman \cite{FRLL} & $270 \chi_g + 0.2829\times 10^6 \chi_g^2 $ & 0.69 & 5006 & $171 \pm 4$ & $2602 \pm 4$ & $2853 \pm 4$ \\
 & $+ 171120\exp(-2.15 \chi_g^2)$ & & & & & \\
\hline
\end{tabular}
\end{table}

\begin{table}
\caption{\label{tab:3} Leading terms of the inverse residence time for binary superpositions of the dimer-oscillator states sketched in Fig. \ref{fig:6}.}
\begin{tabular}{ccccccccc}
\hline
 & $|000\rangle$ & $|010\rangle$ & $|001\rangle$ & $|011\rangle$ & $|100\rangle$ & $|110\rangle$ & $|101\rangle$ & $|111\rangle$ \\
\hline
$|000\rangle$ & $--$ & $2\varepsilon \nu_{01}$ & $2\varepsilon \nu_{01}$ & $4\varepsilon^2 \nu_{01}$ & $8\varepsilon^3 \nu_{01}$ & $2\varepsilon^2 \nu_{01}$ & $2\varepsilon^2 \nu_{01}$ & $4\varepsilon^2 \nu_{01}$ \\
$|010\rangle$ & $2\varepsilon \nu_{01}$ & $--$ & $--$ & $2\varepsilon \nu_{01}$ & $2\varepsilon^2 \nu_{01}$ & $8\varepsilon^3 \nu_{01}$ & $8\varepsilon^4 \nu_{01}$ & $2\varepsilon^2 \nu_{01}$ \\
$|001\rangle$ & $2\varepsilon \nu_{01}$ & $--$ & $--$ & $2\varepsilon \nu_{01}$ & $2\varepsilon^2 \nu_{01}$ & $8\varepsilon^4 \nu_{01}$ & $8\varepsilon^3 \nu_{01}$ & $2\varepsilon^2 \nu_{01}$ \\
$|011\rangle$ & $4\varepsilon^2 \nu_{01}$ & $2\varepsilon \nu_{01}$ & $2\varepsilon \nu_{01}$ & $--$ & $4\varepsilon^2 \nu_{01}$ & $2\varepsilon^2 \nu_{01}$ & $2\varepsilon^2 \nu_{01}$ & $8\varepsilon^3 \nu_{01}$ \\
\hline
\end{tabular}
\end{table}

\clearpage

\section*{list of tables}

\contentsline {table}{\numberline {I}{\ignorespaces Neutron single crystal diffraction data and structure refinement. $\lambda $ = 0.8305 \r A. Space group $P 2_1/c$. $\sigma $(I) limit: 3.00. Refinement on F. The occupation numbers for the proton sites are $\rho _L$ and $\rho _R$. The variance for the last digit is given in parentheses.}}{26}{}
\contentsline {table}{\numberline {II}{\ignorespaces Double-well operators, distances between minima ($\Delta \chi _\xi $), barrier heights ($H_\xi $), and energy levels, for the crystal of the ring deuterated Benzoic acid. $\chi _\xi $ are the symmetry species infrared- ($\chi _u$) or Raman-active ($\chi _g$). The oscillator mass consistent with $\Delta \chi _\xi $ is 1 amu. *: observed with inelastic neutron scattering after Ref. [\ref@citealpnum {PFJHT}]. }}{27}{}
\contentsline {table}{\numberline {III}{\ignorespaces Leading terms of the inverse residence time for binary superpositions of the dimer-oscillator states sketched in Fig. 6{}{}{}\hbox {}.}}{27}{}

\section*{list of figures}

\contentsline {figure}{\numberline {1}{\ignorespaces Sketch of tautomerism (I) and proton interconversion (II).}}{4}{}
\contentsline {figure}{\numberline {2}{\ignorespaces (Color on line) Temperature laws for the occupancy factor of the less favored proton sites. $\bullet $ with error bars: experimental points (this work). $\times $: experimental points after Ref. [\ref@citealpnum {WSF1}]. Blue dot dashed with bars: SOF due to tautomerism only. Solid plus grey zone: Tautomerism plus interconversion without the intermediate states (14{}{}{}\hbox {}). Red dashed with bars: Tautomerism plus interconversion including the intermediate states (12{}{}{}\hbox {}). The grey zone and the bars represent dispersions of the theoretical curves due to uncertainties for the parameters. Curves A and B are analogous to (14{}{}{}\hbox {}) and (12{}{}{}\hbox {}), respectively, for a single tautomer state (see text). }}{7}{}
\contentsline {figure}{\numberline {3}{\ignorespaces CO bond lengths as a function of the occupancy of the less favored proton sites. $\blacksquare $ and $\blacklozenge $ with error bars: experimental points (this work). $\square $ and $\lozenge $: after Ref. [\ref@citealpnum {WSF2}]. The straight lines are guides for the eyes.}}{8}{}
\contentsline {figure}{\numberline {4}{\ignorespaces CO bond lengths as a function of the temperature. $\blacksquare $ and $\blacklozenge $ with error bars: experimental points from Table I{}{}{}\hbox {}. Solid curves were calculated with (4{}{}{}\hbox {}) and $R_\mathrm {C=O} = 1.230(5)$ \r A, $R_\mathrm {C-OH} = 1.318(5)$ \r A, $\Delta E = (78 \pm 20)$ K$^{-1}$. The grey zones represent the statistical dispersion. The dashed curves (5{}{}{}\hbox {}) hold for the coexistence of both tautomers at elevated temperature [\ref@citealpnum {Fil7}].}}{9}{}
\contentsline {figure}{\numberline {5}{\ignorespaces Perspective view of the potential operator for the hydrogen bonding protons in the crystal of benzoic acid. $\chi _{u}$ and $\chi _{g}$ are the symmetry species of the OH stretching mode unveiled by vibrational spectra. }}{12}{}
\contentsline {figure}{\numberline {6}{\ignorespaces Schematic view of the wave functions of the unit cell $j$ for the hydrogen bonding protons in the crystal of benzoic acid. $\chi _{uj}$ and $\chi _{gj}$ are the symmetry species of the unit cell unveiled by vibrational spectra. For the sake of clarity, the weak component of the wave function in one-dimension is multiplied by a factor of 5. The quantum numbers are $l,n_u,n_g$, respectively. }}{13}{}
\contentsline {figure}{\numberline {7}{\ignorespaces Comparison of the experimental temperature law of BA (solid line) (18{}{}{}\hbox {}) with the quantum theory (dot-dashed) (19{}{}{}\hbox {}): $\nu _{01} = 5.16 \times 10^{12} s^{-1}$, $\varepsilon = 0.018$, $\tau _{Q}^{-1} = (2.3\pm 1)10^8 \mathop {\mathgroup \symoperators coth}\nolimits [(39\pm 10)/T] + (6.77\pm 1.15)10^{9}\{\mathop {\mathgroup \symoperators exp}\nolimits [-(89\pm 13)/T] + \mathop {\mathgroup \symoperators exp}\nolimits [-(167\pm 13)/T]\} + (7.2\pm 1.2)10^{11} \mathop {\mathgroup \symoperators exp}\nolimits [-(385\pm 9)/T]$, in proton per second units. The grey zone features the dispersion of the theoretical curve due to uncertainties. The white curve was obtained with $\varepsilon = 0.016$. }}{19}{}

\end{document}